\documentclass[reprint, superscriptaddress, secnumarabic, amssymb, nobibnotes, aps, prl]{revtex4-1}

\usepackage{graphicx}
\usepackage{epstopdf}
\usepackage[T1]{fontenc}
\usepackage[latin9]{inputenc}
\usepackage{amsbsy}
\usepackage{gensymb}
\usepackage[T1]{fontenc}
\usepackage[latin9]{inputenc}
\usepackage{amsmath}
\usepackage{amssymb}
\usepackage{bbm}
\usepackage{braket}
\usepackage{xcolor}
\allowdisplaybreaks
\usepackage{graphicx}
\usepackage[colorlinks=true]{hyperref}  
\hypersetup{
    bookmarks=true,         
    unicode=false,          
    pdftoolbar=true,        
    pdfmenubar=true,        
    pdffitwindow=false,     
    pdfstartview={FitH},    
    pdftitle={RhSe$_2$ and RhSeTe},    
    pdfauthor={a},     
    pdfsubject={},   
    pdfcreator={},   
    pdfproducer={}, 
    pdfkeywords={} {} {}, 
    pdfnewwindow=true,      
    colorlinks=true,       
    linkcolor=blue, 
    citecolor=blue,        
    filecolor=magenta,      
    urlcolor=blue           
} 
\usepackage[normalem]{ulem}

\newcommand{\equref}[1]{Eq.~(\ref{#1})}

\newcommand{\figref}[1]{Fig.~\ref{#1}}

\newcommand{\tableref}[1]{Table~\ref{#1}}

\begin{document}
\title{\textrm{Superconducting Properties of Topological Semimetal 1$T$-RhSeTe}}
\author{C. Patra}
\affiliation{Department of Physics, Indian Institute of Science Education and Research Bhopal, Bhopal, 462066, India}
\author{T. Agarwal}
\affiliation{Department of Physics, Indian Institute of Science Education and Research Bhopal, Bhopal, 462066, India}
\author{Arushi}
\affiliation{Department of Physics, Indian Institute of Science Education and Research Bhopal, Bhopal, 462066, India}
\author{P. Manna}
\affiliation{Department of Physics, Indian Institute of Science Education and Research Bhopal, Bhopal, 462066, India}
\author{N. Bhatt}
\affiliation{Department of Physics, Indian Institute of Science Education and Research Bhopal, Bhopal, 462066, India}
\author{R. S. Singh}
\affiliation{Department of Physics, Indian Institute of Science Education and Research Bhopal, Bhopal, 462066, India}
\author{R. P. Singh}
\email[]{rpsingh@iiserb.ac.in}
\affiliation{Department of Physics, Indian Institute of Science Education and Research Bhopal, Bhopal, 462066, India}

\begin{abstract}
\begin{flushleft}
\end{flushleft}
Platinum-group transition-metal dichalcogenides have emerged as a subject of considerable interest in condensed matter physics due to their remarkable topological properties and unconventional superconducting behavior. In this study, we report the synthesis and superconducting characteristics of a new Dirac-type topological semimetallic compound 1$T$-RhSeTe. It shows type-II superconductivity with a superconducting transition temperature of 4.72 K and a high upper critical field. The coexistence of superconductivity and topological properties makes it a prime candidate for hosting topological superconductivity.

\end{abstract}
\maketitle

\section{INTRODUCTION}

Topological semimetals have emerged as a new study area in quantum materials research. Unlike topological or crystalline insulators, semimetals have minimal overlaps between their bulk conduction and valence bands. This unique characteristic distinguishes them from insulating materials. Topological semimetals are classified into types, including topological Dirac and Weyl semimetals. More recently, this classification has been extended to include nodal lines and non-symmorphic semimetals \cite{high_Tc_Sc,NiS2_mott_insulator,topology_insulator,Dirac_A3Bi,VAl3_dirac,Dirac_Cd3As2, inversion_symmetry_dirac,symetery_dirac_weyl,symmerty_lokking}. The exploration of topological semimetals has yielded remarkable discoveries, such as topological superconductivity, negative magnetoresistance, chiral magnetic effects, and the quantum anomalous Hall effect. These properties hold significant potential to develop cutting-edge technologies \cite{axial_signature,transport_dirac,transport_Bi2_family}. Despite these exciting findings, the main focus of research on topological semimetals has been centered around the discovery of new Dirac and Weyl semimetals, as well as the intricate study of their band topology \cite{VAl3_dirac, Dirac_Cd3As2, TaAs, wte2, MoTe2_weyl}. At the same time, superconducting topological semimetals remain largely unexplored territory despite their potential for direct application in quantum computing.\\

Transition-metal dichalcogenides (TMDs) offer a promising avenue to achieve topological superconductivity \cite{Theory_MoTe2, structure_stable, rhIrTe2_thermal}. These materials can exist in various crystal structures, including hexagonal (2$H$ and 4$H$), rhombohedral (3$R$), trigonal (1$T$), cubic pyrite-type, and orthorhombic forms. The trigonal structure, in particular, is known to host bulk Dirac type-II fermions and topological surface states \cite{bulk_dirac_cone}. Notably, the 1$T$ phases of NiTe$_2$, PdTe$_2$, PtSe$_2$, PtTe$_2$ and IrTe$_2$ exhibit a tilted Dirac cone in the vicinity of the Fermi level \cite{NiTe2_dirac, PtTe2_dirac, PdTe2_dirac, IrTe2_nano_flake, PtSe2_dirac}. Among these compounds, NiTe$_2$ and IrTe$_2$ possess Dirac points very near the Fermi level ($\sim$ 80 meV and 0.15 eV, respectively) \cite{IrTe2_dirac_energy}, which can be fine-tuned through application of strain, pressure or chemical doping. This tunability has been shown to induce or enhance superconductivity in these materials \cite{Pt_doped_sc_IrTe2, IrTe2_nano_flake, spectroscopy_IrTe2_Pt_doped, ARPES_IrTe2}. The ability to control the positions of Dirac points in the 1$T$ phase of chalcogenides makes them very promising for realizing topological superconductivity \cite{dirac_cone_IrTe2}.  Therefore, searching for new 1$T$ topological semimetals is of utmost importance to advance research in the field. \\

Rhodium-based dichalcogenide compounds have been relatively underexplored, tending to stabilize in 1$T$ or pyrite-type crystal structures \cite{structure_stable,rhIrTe2_thermal,layered_structure_irte2,Tc_IrTe2,first_principle_IrTe2,RhSe2_IrSe2}. RhSe$_2$ typically adopts a pyrite-type structure, while RhTe$_2$ can exhibit both pyrite-type and 1T structures, depending on the Rh/Te ratio. Interestingly, in the pyrite-type structure, RhSe$_2$ and RhTe$_2$ exhibit superconductivity with transition temperatures of 6 K and 1.5 K, respectively \cite{RhSe2_sc_first,RhSe2_paper_sc,RhTe2_first_paper,RhTe_phase}. Investigations on Rh substitution at the Ir site in these compounds demonstrate the formation of a 1$T$ structure with a maximum transition temperature ($T_c$) of ~2.6 K when the Rh content ranges from 0 to 0.3 in standard solid-state synthesis \cite{RhSe2_IrSe2}. In the Ir${1-x}$Rh$_x$Se$_2$ system, high-pressure synthesis leads to the formation of a pyrite-type structure, resulting in a dome-shaped variation in $T_c$, with the highest $T_c$ of 9.6 K achieved at $x = 0.36$ \cite{RhSe2_IrSe2}. These findings suggest that disorder/charge transfer can induce or enhance superconductivity in 1$T$/Pyrite phases of RhTe$_2$.\\

\begin{figure*}
	\includegraphics[width=2.05\columnwidth]{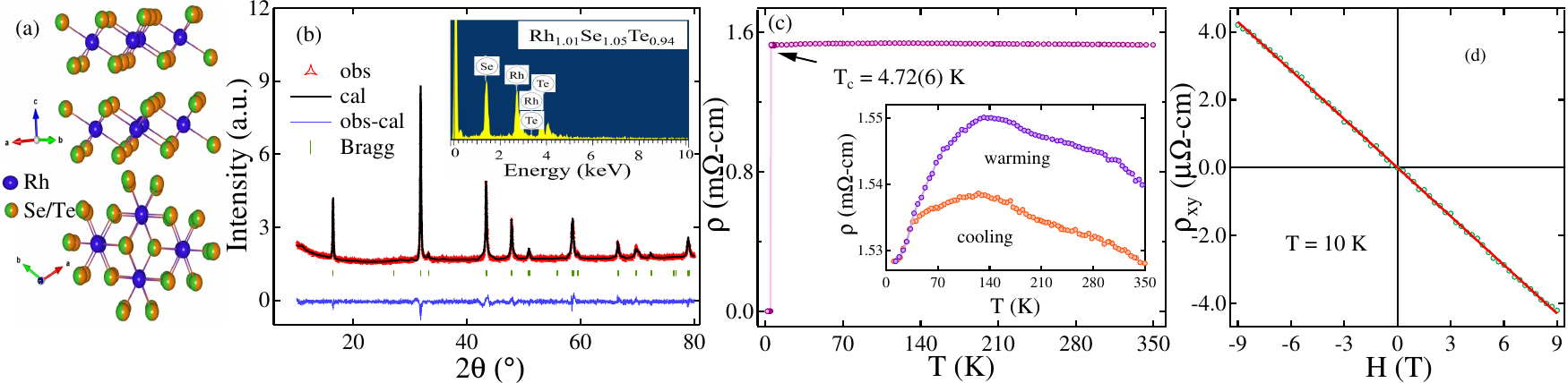}
	\caption{\label{XRD} (a) shows the crystal structure of RhSeTe. (b) Powder XRD pattern under ambient conditions for RhSeTe. The structural refinement was performed using the space group of $P\Bar{3}m1$. The inset shows the EDS scan that detects Rh, Se, and Te. (c) Resistivity shows the superconducting transition at 4.72(6) K. The inset reveals hysteresis in the normal-state resistivity during heating and cooling cycles. (d) The Hall resistivity at 10 K exhibits a linear behavior to the applied magnetic field variation ($\pm$ 9 T).}
\end{figure*}

Although RhTe2 can stabilize in the 1T structure, superconductivity has not been reported in this phase \cite{rhIrTe2_thermal,RhSe2_sc_first}. Sametime, coexistence of pyrite and the 1T structure presents a challenge in obtaining a pure 1T phase of RhTe2. However, in our current study, we achieved a pristine 1T-phase of RhSeTe by substituting selenium for tellurium in RhTe2 \cite{PdTeSe_dirac}. Notably, this RhSeTe exhibits a superconducting transition below 4.72 K. Resistivity measurements reveal significant hysteresis, indicating the presence of a charge density wave. Density functional theory (DFT) calculations suggest the presence of a type-II Dirac cone located  ~82 meV above the Fermi level in RhSeTe. The electronic band structure closely resembles the isostructural topological semimetal 1$T$-NiTe$_2$. Phonon band structure calculations indicate unstable phonon modes away from the zone center, further supporting the potential for charge density wave formation, possibly linked to the observed resistivity hysteresis.\\

\section{EXPERIMENTAL METHODS}

A polycrystalline sample of RhSeTe was prepared by the standard solid-state reaction method. The stoichiometric mixtures of Rh (4N), Se (5N) and Te (5N) were thoroughly grounded, pelletized, and sealed in an evacuated quartz tube. The sample was heated to 1100\degree C for a week, followed by quenching with ice water to avoid the formation of unwanted impure phases. The purity of the RhSeTe phase was determined using powder X-ray diffraction (XRD) on a PANalytical diffractometer equipped with Cu$K_{\alpha}$ radiation ($\lambda$ = 1.54056 $\text{\AA}$). Furthermore, magnetization measurements were performed using a quantum interference device (MPMS3, Quantum Design). The resistivity and specific heat of the sample were measured in the Physical Property Measurement System (PPMS) using the conventional four-probe AC technique and the two-tau relaxation method, respectively.

\section{RESULTS AND DISCUSSION}

\subsection{Sample characterization}

The crystal structure of 1$T$ -RhSeTe from side and top views is represented in \figref{XRD}(a). The phase purity of the sample and the crystal structure were determined from the Rietveld refinement of the powder XRD pattern using FullProf software, shown in \figref{XRD}(b). It confirms that RhSeTe adapts the trigonal structure with space group: $P\Bar{3}m1$ (space group no. 164) \cite{structure}. The extracted lattice parameters are a = b = 3.798(2)$\text{\AA}$, and c = 5.389(5)$\text{\AA}$ with $\gamma$ = 120$\degree$. EDS measurements conducted in different regions revealed an average elemental concentration, Rh$_{1.01}$Se$_{1.05}$Te$_{0.94}$, as depicted in the inset of \figref{XRD}(b). All parameters, including space group, atomic, and Wyckoff positions, acquired from refinement are summarized in \tableref{elec_XRD}.

\begin{table}[h!]
\caption{Parameters obtained from Rietveld refinement.}
\label{elec_XRD}
\begin{center}
\begin{tabular*}{1.0\columnwidth}{l@{\extracolsep{\fill}}lcccc}\hline\hline
&  & Structure & Trigonal &   &\\
& & space-group & $P\Bar{3}m1$ (164) & &\\
& & lattice parameters & & &\\
\hline
& &$a=b$ ($\text{\AA}$) & 3.798(2) & \\
& & $c$ ($\text{\AA}$) & 5.389(5) &\\
&Atom & Wyckoff position & x & y & z \\
\hline
&Rh1 & 1a & 0 & 0 & 0  \\
&Se1 & 2d & 0.333 & 0.667 & 0.256  \\
&Te1 & 2d & 0.333 & 0.667 & 0.256  \\
\hline\hline
\end{tabular*}
\par\medskip\footnotesize
\end{center}
\end{table}

\subsection{Transport measurement}

The temperature dependence of the electrical resistivity $\rho(T)$ was measured in a zero applied magnetic field over the range of 350 to 1.9 K and is represented in \figref{XRD}(c). The low-temperature resistivity region is depicted, and it abruptly reaches zero at 4.72(6) K, confirming the occurrence of a superconducting transition \cite{structure,RhTe2_first_paper}. The resistivity of 1$T$-RhSeTe in the normal state demonstrates a negligible temperature dependence. In contrast, for 1$T$-RhTe$_2$, the resistivity exhibits a metallic-type variation with temperature \cite{rhIrTe2_thermal}. This distinction can potentially be attributed to the high disorder introduced by the isoelectronic selenium atom within the system. Interestingly, the resistivity of 1$T$-RhSeTe shows a notable behavior with a negative temperature coefficient above approximately 140 K, as shown in the inset of \figref{XRD}(c). Moreover, a large hysteresis is observed between the temperature-dependent resistivity heating and cooling cycles. The hysteresis and negative temperature coefficient may originate from a CDW transition or dimerization. A similar hysteresis behaviour is observed in 1$T$-IrTe$_2$ at 260 K, originating by stripe charge ordering accompanied by the first-order structural transition involving dimerization or depolymerization \cite{layered_rh_doped_IrTe2,spin_orbit_coupling_IeTe2}.

Furthermore, to extract the carrier concentration of RhSeTe, the field-dependent Hall resistivity $\rho_{xy}(H)$ at 10 K was measured as shown in \figref{XRD} (d). $\rho_{xy}(H)$ is well described by a straight-line fit and provides the Hall coefficient $R_H$ = -4.77(4)$\times$ 10$^{-8}$ $\ohm$-cm/T for RhSeTe. The negative sign indicates the presence of electron carrier concentration in the system, and using the relation $R_H = 1/{ne}$ (where e is the electronic charge), the calculated carrier concentration n = 1.31(4)$\times$ 10$^{22}$ cm$^{-3}$ at 10 K, which is of the same order as that reported for 1$T$-RhTe$_{2}$ \cite{rhIrTe2_thermal}.



\subsection{Magnetization}

\begin{figure}
	\includegraphics[width=1.0\columnwidth]{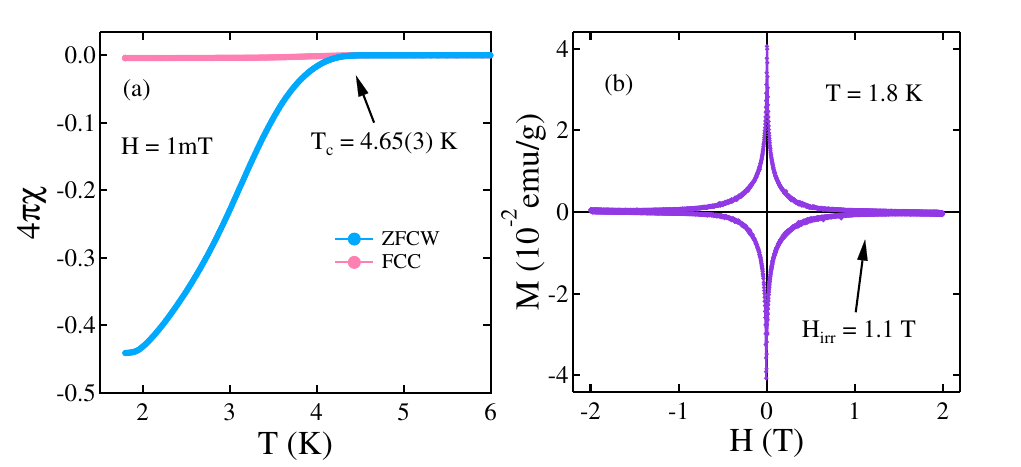}
	\caption{\label{tc} (a) Magnetization data at ZFCW-FCC mode showing the superconducting transition at 4.65(3) K for RhSeTe. (h) Magnetization vs magnetic field at 1.8 K indicates type-II superconducting behaviour and H$_{irr}$ estimated 1.1 T at 1.8 K.}
\end{figure}

Magnetic susceptibility was measured in zero-field-cooled-warming mode (ZFCW) and field-cooled-cooling mode (FCC) under 1 mT magnetic field for bulk 1$T$-RhSeTe. The onset of diamagnetism confirmed the superconductivity in RhSeTe with the appearance of a superconducting transition temperature $T_c^{onset}$ = 4.65(3) K, as shown in \figref{tc}(a). \figref{tc}(b) displays the full range magnetization with field ($M$-$H$) at $T$ = 1.8 K, indicating type-II superconducting behaviour of RhSeTe, and magnetization is irreversible below a field $H_{irr}$ = 1.1 T.\\

\begin{figure}
\includegraphics[width=0.96\columnwidth]{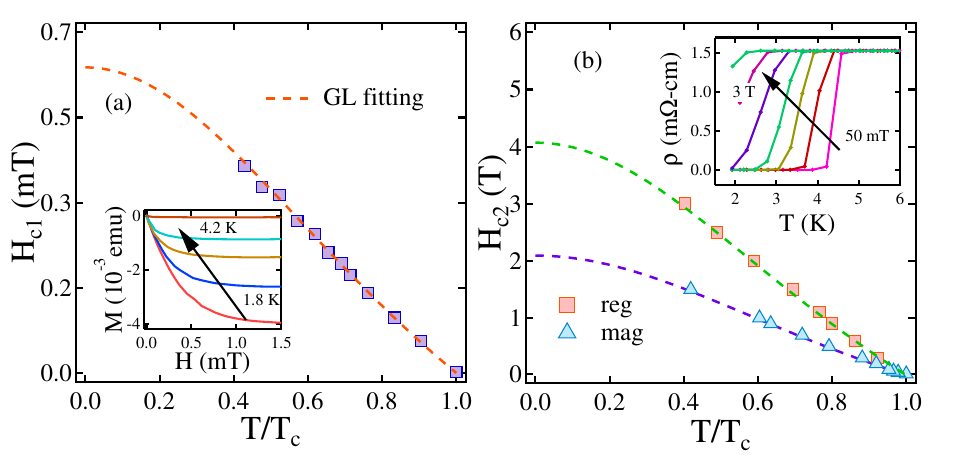}
	\caption{\label{hc1_hc2} (a) The temperature variation of $H_{c1}$ for RhSeTe displays the value of lower critical field 0.63(1) mT and the inset shows low field magnetization curves as a function of applied magnetic field for different temperatures. (b) $H_{c2}(T)$ data from magnetization and resistivity fitted with the Ginzburg-Landau model for RhSeTe. The inset illustrates the temperature variation of the resistivity in different fields up to 3 T.}
\end{figure}

The lower critical field, $H_{c1}(0)$, is extracted from the low field magnetization curves as a function of the magnetic field (0-15 mT) at different constant temperatures, which is shown in the inset of \figref{hc1_hc2}(a). The points of $H_{c1}(T)$ are deduced from the linear deviation of the Meissner line from the initial slope of $M$-$H$ curves. The value of $H_{c1}(0)$ = 0.63(1) mT was depicted from the fit using the Ginzburg-Landau equation,

\begin{equation}\label{eqn3}
    H_{c1}(T) = H_{c1}(0)\left[1-\left(\frac{T}{T_c}\right)^2\right]
\end{equation}
and it is presented in \figref{hc1_hc2}(a). The upper critical field, $H_{c2}$(0), is calculated from the temperature variation of the resistivity and magnetization measurements taken at different applied magnetic fields. $T_{c}$ shifts towards the lower temperature with an increasing magnetic field. The inset of \figref{hc1_hc2}(b) shows the temperature-dependent resistivity curves for different fields. From magnetization measurements, $T_{c}$ is considered the onset of the diamagnetic signal.  In the case of resistivity curves, the determination of $T_{c}$ is based on the 90\% value of the normal state resistivity. The temperature variation of $H_{c2}(T)$ was well described using the Ginzburg-Landau equation (\equref{eqn4}),

\begin{equation}\label{eqn4}
   H_{c2}(T) = H_{c2}(0) \left[\frac{1-\left(\frac{T}{T_c}\right)^2}{1+\left(\frac{T}{T_c}\right)^2}\right].
\end{equation}

and yielded $H_{c2}^{mag,\rho}$(0) as 2.09(1) T and 4.06(7) T respectively, shown in \figref{hc1_hc2}(b). Iridium-based dichalcogenides, including pyrite- or trigonal-typeIr$_{0.93}$Te$_2$ shows a superconducting transition temperature of 4.7 K, accompanied by an upper critical field value of 4.58 T \cite{Tc_IrTe2}.\\

The fundamental superconducting parameters, such as coherence length, penetration depth, and thermodynamic critical field, are estimated by using the values of $H_{c2}^{\rho}(0)$ and $H_{c1}(0)$. The Ginzburg-Landau coherence length, $\xi_{GL}(0)$ and the penetration depth, $\lambda_{GL}(0)$ are calculated by using the following equations:

\begin{equation}\label{eqn5}
    H_{c2}^{\rho}(0) = \frac{\Phi_0}{2\pi\xi_{GL}^2}
\end{equation}

\begin{equation}\label{eqn6}
    H_{c1}(0) = \frac{\Phi_0}{4 \pi \lambda^2_{GL}(0)}\left[ln\left(\frac{\lambda_{GL}(0)}{\xi_{GL}(0)}\right) + 0.12\right]
\end{equation}

where $\Phi_0$ (= 2.07$\times10^{-15}$ Tm$^2$) is the magnetic flux quantum. The extracted values of $\xi_{GL}(0)$ and $\lambda_{GL}(0)$ for RhSeTe are 90.1(8) $\text{\AA}$ and 1138(9) nm, respectively. Furthermore, the values of $\xi_{GL}(0)$ and $\lambda_{GL}(0)$ were used to calculate the Ginzburg-Landau parameter $\kappa_{GL}$ = $\frac{\lambda_{GL}(0)}{\xi_{GL}(0)}$ = 126(2). The value of $\kappa_{GL}$ is greater than $\frac{1}{\sqrt{2}}$, indicating the type II superconducting behaviour of the sample. The thermodynamic critical field $H_c$ is obtained from the relation $H_{c1}(0) H_{c2}(0) = H^2_cln{\kappa_{GL}}$ and provides the value $H_c$ = 22 (1) mT.\\

Type-II superconductors exhibit two mechanisms for breaking Cooper pairs in a magnetic field: the orbital effect and the Pauli paramagnetic effect, or Zeeman splitting, as described in \cite{pauli_limit}. In the case of orbital pair breaking, the increased kinetic energy causes the pairs to break. On the contrary, in the Pauli paramagnetic limiting effect, the spins of the Cooper pair attempt to align in the direction of the magnetic field, leading to a difference in energy between the electrons and the consequent breaking of the Cooper pair \cite{WHH_model}. The Pauli paramagnetic limit is expressed as $H_{c2}^P(0) = 1.86T_c$ = 8.78 T for RhSeTe. The orbital pair-breaking limit is given by the Werthamer-Helfand-Hohenberg (WHH) expression:

\begin{equation}\label{eqn8}
    H_{c2}^{orbital} = -\alpha T_c \frac{dH_{c2}(T)}{dT}.
\end{equation}

The initial slope $\frac{dH_{c2}(T)}{dT}$ near $T = T_c$ is calculated to be -1.03(5) T/K. When considering $\alpha$ = 0.693 for the dirty limit, the $H_{c2}^{orbital}$ value is determined to be 3.37 T. Moreover, the Maki parameter \cite{maki_parameter} ($\alpha_M = \frac{\sqrt{2}H_{c2}^{orbital}}{H_{c2}^P}$) is 0.54. The value of the Maki parameter is less than unity, indicating that the influence of the paramagnetic effect is negligibly small in RhSeTe in breaking Cooper pairs under the applied magnetic field.

\subsection{Specific heat}

\begin{figure}
	\includegraphics[width=0.98\columnwidth]{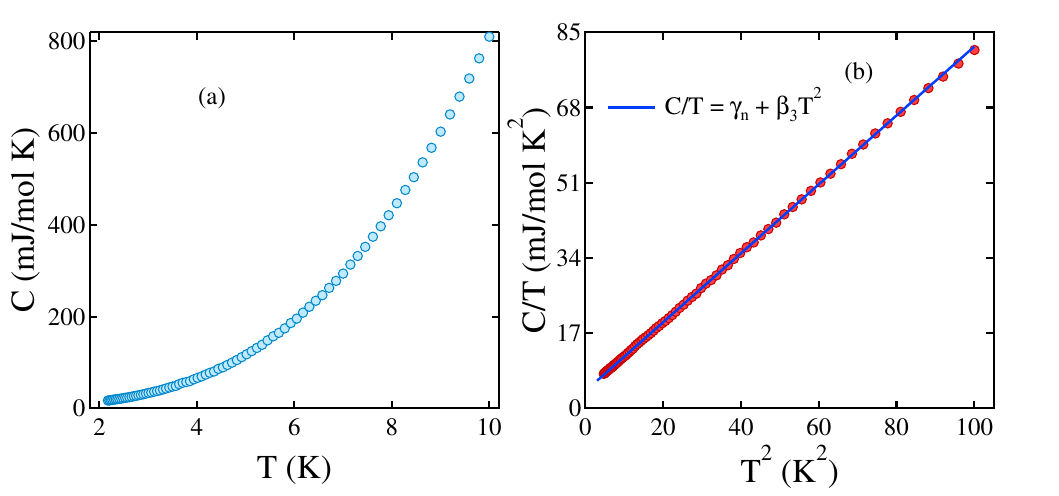}
	\caption{\label{sp} (a) The temperature variation of specific heat. (b) $C/T$ vs $T^2$ representation of specific heat is used to find the electronic and phononic contribution at normal state.}
\end{figure}

Furthermore, the superconducting gap symmetry is explored using the zero-field specific heat measurement. Specific heat data for RhSeTe are collected from 1.9 to 10 K in a zero field, as shown in \figref{sp}(a). The phononic properties of the normal state and the electron density of the states are evaluated in \figref{sp}(b) using \equref{eqn9}.

\begin{equation}\label{eqn9}
    C/T = \gamma_n + \beta_3T^2
\end{equation}

where $\gamma_n$ represents the electronic contribution and $\beta_3$ describes the Debye constant or the phononic contribution. The fitting parameters yield $\gamma_n$ = 3.89(2) mJ/mol-K$^2$ , and $\beta_3$ = 0.78(3) mJ/mol-K$^4$. The specific heat data do not show a distinct jump in the transition temperature, possibly due to the presence of significant disorder in the layered structure, a characteristic of doping in polycrystalline materials. However, it is worth noting that the superconducting jump in single-crystal samples may be observable since single crystals generally exhibit less disorder than polycrystalline materials. From the Debye constant $\beta_3$, the Debye temperature $\theta_D$ is estimated to be 195(5) K using \equref{eqn10}. The reported Debye temperatures for Ir$_{0.94-x}$Rh$_x$Se$_2$ \cite{RhSe2_IrSe2} are closer to the value found for RhSeTe.

\begin{equation} \label{eqn10}
    \theta_D = \left(\frac{12\pi^4RN}{5\beta_3}\right)^{\frac{1}{3}}
\end{equation}

where $R$ is the universal gas constant (=8.314 J mol$^{-1}$ K$^{-1}$) and $N$ = 3, the number of atoms per formula unit.\\

The density of state, $D_C(E_F)$, is calculated from the concept of the Sommerfeld coefficient $\gamma_n$ via the relation $\gamma_n = (\frac{\pi^2k_B^2}{3})D_C(E_F)$, where $k_B$ = 1.38 $\times$ $10^{-23}$ J K$^{-1}$. The estimated value $D_C(E_F)$ is 1.65 states eV$^{-1}$ f.u.$^{-1}$. The calculated values of the density of states and Sommerfeld coefficient from specific heat were verified from the theoretical calculations, as discussed in a later section. Also, the dimensionless parameter, the electron-phonon coupling constant ($\lambda_{e-ph}$), provides information about the strength of the electron-phonon interaction in the system. $\lambda_{e-ph}$ is computed using the expression of $\lambda_{e-ph}$ from McMillan's model \cite{coupling} \equref{eqn14}
\begin{equation}\label{eqn14}
    \lambda_{e-ph} = \frac{1.04+\mu^*ln(\theta_D/1.45T_c)}{(1-0.62\mu^*)ln(\theta_D/1.45T_c)-1.04}
\end{equation}
where $\mu^*$ represents the Coulomb repulsion constant and is taken to be 0.13 \cite{coupling}. The extracted value of $\lambda_{e-ph}$ = 0.72 for RhSeTe is slightly higher than that reported for other weakly coupled superconductors. Previous studies on 2$H$-TaSe$_{2-x}$S$_x$ \cite{TaSexSx,tases_2H}, Ir-doped MoTe$_2$ \cite{IrMoTe2}, and Re-doped NiTe$_2$ \cite{Re_doped_NiTe2} have shown that increasing disorder leads to an increase in electron-phonon coupling and, consequently, enhances the superconducting transition temperature. Similarly, it is possible that in 1$T$-RhSeTe, disorder plays a crucial role in moderate enhancement of $\lambda_{e-ph}$ and, in turn, may be responsible for the increase in $T_{c}$, compared to the 1$T$-RhTe$_{2}$ phase where no superconductivity has been found.

\subsection{Electronic properties and the Uemura plot}


To explore the electronic and phononic characteristics of the system, we conducted Density Functional Theory (DFT) calculations. Our investigation involved employing a 2$\times$2$\times$1 supercell of bulk Rh$X_2$ to assess the impact of the alloying of Te and Se at the $X$ site. To simulate the disorder of chalcogen atoms, we utilized the special quasirandom structure method (SQS) as described in Reference \cite{SQS,SQS1}. The calculations were carried out using the projector augmented wave (PAW) method implemented in the Vienna Ab-initio Simulation Package (VASP) \cite{VASP,PAW,PAW1}. An energy cut-off of 500 eV was set for the plane-wave basis. Exchange-correlation functionals were based on the Generalized Gradient Approximation (GGA) following the non-local vdW-DF functional optB88-vdW proposed by Klime et al. \cite{vdW-DF}, and they were employed to fully optimize the lattice structure. During the structural optimization of the supercell, we ensured that atomic forces remained below 0.5 meV/\AA, and we sampled a $\Gamma$-centered 9$\times$9$\times$9 $k$-mesh across the Brillouin zone. The relaxed unit cell lattice parameters ($a$ = $b$ = 3.806 \AA, $c$ = 5.346 \AA) closely matched the values obtained by experimental means.

Furthermore, our DFT calculations incorporated spin-orbit coupling, and we derived the effective band structure using the VASPKIT program \cite{VASPKIT}. To assess the phonon spectrum, we employed a 2$\times$2$\times$2 supercell of the SQS and applied the frozen phonon approach as implemented in the PHONOPY code \cite{PHONOPY}.

 \begin{figure}
	\includegraphics[width=1.0\columnwidth]{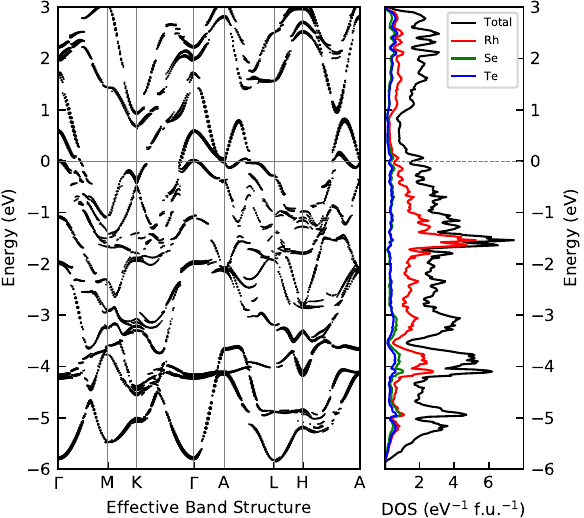}
	\caption{\label{Electron_DOS} Electronic band structure and density of states obtained from DFT calculations including SOC.}
\end{figure}

 \begin{figure}
	\includegraphics[width=1.0\columnwidth]{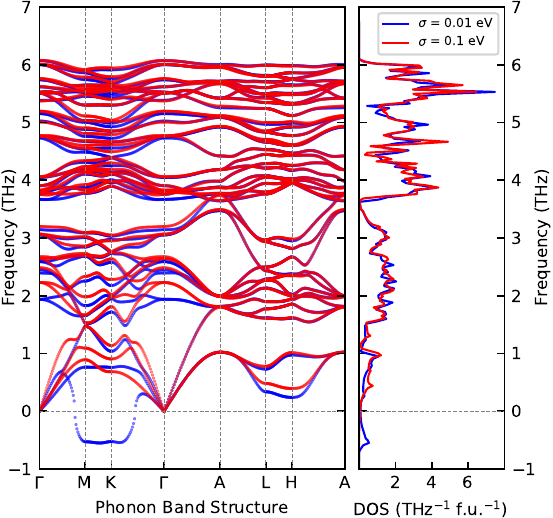}
	\caption{\label{Phonon_DOS} Phononic band structure and density of states.}
\end{figure}

The calculated electronic band structure unfolded in the primitive Brillouin zone, and the density of states is shown in \figref{Electron_DOS}. The band structure is very similar to NiTe$_2$, where two bands that appear around 0.5 eV and 1.3 eV at the $\Gamma$ point in the unoccupied energy range disperse to form the electron pocket centered around the M and K points. These bands cross each other to form a type II Dirac cone while dispersing in the $\Gamma$ - A direction. The energy and momentum position of the Dirac cone was found to be approximately 82 meV above the Fermi level and $k_z$ = 0.394 $c^*$. The density of states at the Fermi level is found to be about 1.61 eV$^{-1}$, leading to a Sommerfeld coefficient of about 3.8 mJ/mol-K$^2$ in excellent agreement with the results obtained from specific heat measurements. The density of states has an increasing trend towards unoccupied, with a maximum of 2.25 eV$^{-1}$ appearing at the energy position of the Dirac crossing. It should be noted that 1$T$-RhTe$_2$ exhibits type II Dirac crossing at approximately 320 meV above Fermi level at $k_z$ = 0.362 $c^*$. The SOC-driven splitting of the Te $p_x$ +$p_y$ band at the $\Gamma$ point is approximately 0.89 eV for RhTe$_2$. This splitting reduces significantly for RhSeTe and becomes approximately 0.57 eV as seen in \figref{Electron_DOS}. The reduction of the SOC strength in the case of RhSeTe leads to the appearance of a Dirac crossing in the close vicinity of the Fermi level.

\figref{Phonon_DOS} shows the calculated phonon dispersion and phonon density of states obtained for the SQS. Phonon dispersion, calculated with an electronic broadening ($\sigma$) of 0.01 eV, exhibits imaginary phonon frequencies along the $\Gamma$-M and $\Gamma$-K directions suggesting structural instability. These unstable acoustic phonons become stable with a larger electronic broadening of 0.1 eV, as shown in \figref{Phonon_DOS}. The signature of a soft-phonon mode can also be seen at the boundary of the Brillouin at the L and H points. It is to note here that a variety of structural distortion and/or charge density orders are observed in various TMDs, more specifically in 1T type TMDs leading to 1T$^\prime$, 1T$^{\prime \prime}$ and 1T$^{\prime \prime \prime}$ structures. The observation of imaginary phonon frequencies away from the zone center and its stabilization with increased electronic broadening suggests a charge density wave formation in this system similar to other TMDs, e.g., TaS$_2$, NbSe$_2$, etc. The observed hysteresis in the cooling and warming cycles in the broad temperature range of the resistivity measurement may be related to these unstable/soft phonon modes.     \\


To quantify the electronic parameters for RhSeTe, such as effective mass, mean free path, and Fermi velocity, first, the Fermi wave vector, $k_F$, is estimated from the carrier concentration deduced from Hall measurement using the relation $k_F = (3\pi^2n)^{1/3}$ \cite{book_parameter}. Here $n$ = 1.31 $\times$ 10$^{22}$ cm$^{-3}$, which yields $k_F$ = 0.73 (1) $\times$ 10$^{10}$ $\text{\AA}^{-1}$. The Sommerfeld coefficient $\gamma_n$ and $k_F$ are used to calculate the effective mass, $m^* = (\hbar k_F)^2\gamma_n/\pi^2nk_B$, where $k_B$ is the Boltzmann constant, and $m^*$ becomes 2.45(3)m$_e$. The mean free path is defined as $l_e = v_F\tau$, where $\tau$ the scattering time given by $\tau=(ne^2\rho_0/m^*)^{-1}$ and $v_F = \hbar k_F/m^*$, $\rho_0$ is the residual resistivity. Thus, the values of $\rho_0$, $\tau$, and $m^*$ provide the value of the mean free path, $l_e$ = 1.50(2) $\text{\AA}$. The coherence length $\xi_0$ = 0.18$\hbar v_F$/$k_B T_c$ is calculated as 1006(2) $\text{\AA}$. The $\frac{\xi_0}{l_e}$ ratio = 670, indicating that RhSeTe is in the dirty limit.

\begin{figure}
	\includegraphics[width=1.0\columnwidth]{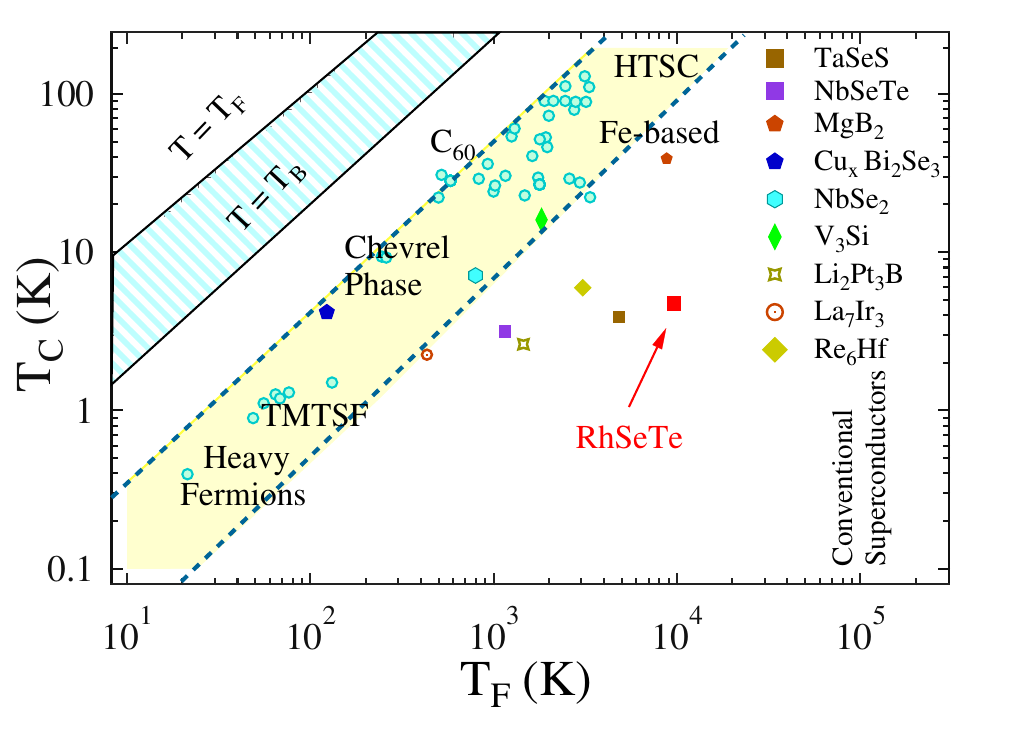}
	\caption{\label{Uemura} The Uemura plot represents the superconducting transition temperature versus the Fermi temperature for different superconducting families. The region between two dotted blue lines shows the unconventional band of superconductors. RhSeTe lies close to the unconventional band.}
\end{figure}

Uemura et al. proposed that superconducting materials can be considered conventional or unconventional based on the $T_c/T_F$ ratio \cite{uemra_plot_2}.  Chevrel phases, heavy fermions, Fe-based superconductors, and high $T_c$ superconductors are placed in the unconventional category as the $T_c/T_F$ ratio falls within 0.01 $\le$T$_c$/T${_F}\le$0.1. To classify RhSeTe, the Fermi temperature $T_F$ has been evaluated by the following equation:
\begin{equation}
    k_BT_F = \frac{\hbar^2}{2} (3\pi^2)^{2/3}\frac{n^{2/3}}{m^*}
\end{equation}
where $n$ = 1.31 $\times$ 10$^{28}$ m$^{-3}$ and m$^*$ = 2.45m$_e$ for the sample. The calculated Fermi temperature for RhSeTe is 9621(44) K (as shown in \figref{Uemura}). The obtained ratio of $T_c/T_F$ = 0.00049 places RhSeTe close to the unconventional band, where there are a few other unconventional superconductors \cite{tases_2H,NbSeTe_uenra,1st_uemra,uemra_equation}.
\begin{table}[h!]
\caption{Parameters in the normal and superconducting state of RhSeTe}
\label{Tab:electronic_properties}
\begin{center}
\begin{tabular*}{1.0\columnwidth}{l@{\extracolsep{\fill}}ccc}\hline\hline
Parameter& Unit & RhSeTe\\
\hline
\\[0.5ex]
$T_{c}$&K  & 4.72(6)\\
 $H_{c1}(0)$ & mT &0.63(1) \\
   $H_{c2}^{\rho}(0)$& T &4.06(7) \\
   $H_{c2}^{P}$& T &8.77(9) \\
   $\xi_{GL}(0)$& $\text{\AA}$ &90.08(8) \\
   $\lambda_{GL}(0)$& nm &1138(9) \\
   $\gamma_n$& mJ/mol-K$^2$ &3.89(2) \\
   $\beta_3$& mJ/mol-K$^4$ &0.78(3) \\
   $\theta_D$&K & 195(5)\\
   $\lambda_{e-ph}$&  &0.72(1) \\
   $n$(10 K)& 10$^{22}$cm$^{-3}$ &1.31(4) \\
   $\frac{m^*}{m_e}$&   &2.45(3) \\
   $l_e$& $\text{\AA}$  &1.50(2) \\
   $\xi_0$& $\text{\AA}$  &1006(2) \\
   $\frac{\xi_0}{l_e}$&  &670(4) \\
   $T_F$& K & 9621(44) \\
\\[0.5ex]
\hline\hline
\end{tabular*}
\par\medskip\footnotesize
\end{center}
\end{table}

Our findings reveal that RhSeTe exhibits the features of a type-II Dirac semimetal, characterized by a tilted Dirac cone, and the proximity of the Dirac point to the Fermi surface, indicating the emergence of a topological phase. This property presents an intriguing opportunity, as the inclusion of isoelectronic disorder in the parent compounds RhSe$_2$ and RhTe$_2$ allows fine-tuning of the Dirac point to approximately 82 meV above the Fermi surface. Notably, this proximity of the type-II Dirac point to the Fermi surface, reminiscent of the observations in NiTe$_2$, is of considerable interest.

Moreover, we observe an increase in the electron-phonon coupling constant, suggesting that the introduced disorder triggers superconductivity. Consequently, this disorder-controlled layered TMD system, 1$T$-RhSeTe, serves as a distinctive platform for exploring topological superconductivity.

\section{Conclusion}

The study investigated the superconductivity of 1$T$-RhSeTe at 4.72 K, which offers a unique platform to explore CDW and Dirac/Weyl semi-metals. Substituting Te in RhSe$_2$ resulted in a stabilized trigonal structure instead of the expected cubic pyrite structure under ambient conditions. The results shed light on the origin of superconductivity in Rh-based systems and show that disorder plays a crucial role in determining the structure and enhancement of the upper critical field and other superconducting parameters. The superconducting and thermodynamic parameters are listed in \tableref{Tab:electronic_properties}. Observing insulating behavior above 140 K and hysteresis in resistivity indicates a CDW and dimerization in the system. Furthermore, density functional theory (DFT) calculations suggest the presence of a Dirac type-II node located approximately 82 meV above the Fermi surface, very similar to 1$T$-NiTe$_2$ and instability in phonon spectra related to charge density wave. Since the current study indicates that the manipulation of the Se/Te ratio can control the location of the Dirac point on the Fermi surface, it can offer a novel perspective on understanding the interplay between the position of Dirac points and superconductivity in topological semimetallic compounds.

\section{Acknowledgments} R.~P.~S. acknowledges the SERB, Government of India, for the Core Research Grant CRG/2019/001028.

\end{document}